# On undrained test using Rowe's relation and Incremental Modelling: Generalisation of the notion of Characteristic State

## P. Evesque

Lab MSSMat, UMR 8579 CNRS, Ecole Centrale Paris
92295 CHATENAY-MALABRY, France, e-mail evesque@mssmat.ecp.fr

**Abstract:**
*It is recalled that stress-strain incremental modelling is a common feature of most theoretical description of the mechanical behaviour of granular material. An other commonly accepted characteristics of the mechanical behaviour of granular material is the Rowe's relation which links the dilatancy $K= -\partial\varepsilon_v/\partial\varepsilon_1$ to the stress ratio $\sigma_1/\sigma_3$ during a $\sigma_2=\sigma_3=c^{ste}$ test, i.e. $\sigma_1/\sigma_3=(1+M)(1+K)$. Using the incremental modelling, this law shall be interpreted as a pseudo-Poisson coefficient. We combine these two features to solve the problem of an axial compression under undrained condition. We demonstrate that the sample is submitted to a bifurcation of the transcritical type when it reaches the q=Mp line. This allows to extend the notion of the characteristic state introduced by Luong to other situations and to anisotropic systems. We show also that these undrained tests are quite appropriate to study the characteristic-state behaviour.*
_______________________________________________________________________

The aim of this paper is to point out that using the Rowe's equation [1], which is one of the most classical experimental result on granular materials and on other soils, one can derive the main features of an undrained test by applying a simple incremental modelling. In a first part we recall briefly the Rowe's law of dilatancy, the incremental modelling (or the hypo-elastic modelling) [2] and what their combination implies. More details on this description can be found in [3]. The second part describes what one can expect from this modelling when one applies it to an undrained test (i.e. $v=c^{ste}$ test); the concept of characteristic state is discussed and extended in the light of the incremental modelling. In the last part, few remarks are made concerning the difficulty to apply a mean field treatment and its danger and concerning the existence of a well defined energy dissipation function for granular matter; these points will be discussed together with their implication on the validity of the incremental approach.

### 1. Incremental modelling using Rowe's relation: the basis

***Rowe's law of dilatancy:*** Rowe [1] has proposed some relationship which relates the stress field ($\sigma_1$, $\sigma_2=\sigma_3$) in an axi-symmetric triaxial test and the dilatancy K of the material; K is defined as $K=-\delta\varepsilon_v/\delta\varepsilon_1$, where $\varepsilon_v$ is the volume deformation and $\varepsilon_1$ is the axial one. His approach consists in considering regular arrays of cylinders and in analysing the stress field which is required to impose the sliding of a given row of cylinders, taking into account the stress field and the friction at contact points. As the orientation of the row is in general not parallel to the local surfaces of contact, it results that i) the stress field required to impose the row motion does not correspond to that one of the sliding along the row direction, but does correspond to a sliding parallel to the surface of contacts, and that ii) the misfit of orientation between the row direction and the contact surfaces leads to a volume change. As these two mechanisms are linked together they impose some equation which links them together; this is the Rowe's equation.   Rowe has demonstrated that this relation can be written in such a way that it does not depend on the nature of the 2d lattice (triangular, square,…):

$$\sigma_1/\sigma_3 = (1+K)\tan^2(\pi/4+\varphi/2) = (1+M)(1+K) \quad (1)$$

with φ being the friction angle. So, Rowe has generalised his results and concluded that this relation should be valid whatever the lattice, even when it is 3d and disordered. Experimental tests have been performed to check the validity of the relation using axi-symmetric triaxial test at $\sigma_2 = \sigma_3 = c^{ste}$ and relatively good agreement has been found [1,4].

***Hypoelastic modelling of plastic behaviour:*** let us first consider a sand sample submitted to a given stress field ($\sigma_1$, $\sigma_2 = \sigma_3$) and let us consider any incremental deformation $\underline{\delta\varepsilon}=(\delta\varepsilon_1, \delta\varepsilon_2, \delta\varepsilon_3)$; it is most likely that one can force the sample to deform according to this path by applying an increment of stress $\underline{\delta\sigma}=(\delta\sigma_1, \delta\sigma_2, \delta\sigma_3)$; hence, any set of infinitely small deformation $(\delta\varepsilon_1, \delta\varepsilon_2, \delta\varepsilon_3)$ is possible; in other words, the evolution of this sample is governed by an incremental law which relates the increment of stress tensor $\underline{\delta\sigma}$ to the increment of strain tensor $\underline{\delta\varepsilon}$ so that one can write a relation of the kind g($\underline{\delta\varepsilon}$, $\underline{\delta\sigma}$, $\underline{\sigma}$)=0; furthermore, as the evolution of a sand sample does not depend only on its present stress field but also on its story, g shall be a function which is story dependent.

In order to characterise g, it is necessary to introduce the objectivity principle, which states that the response $\underline{\delta\varepsilon}$ shall be unique for a given increment $\underline{\delta\sigma}$ applied to a given sample under specified condition; so one shall be able to write $\underline{\delta\varepsilon}$ in the form $\underline{\delta\varepsilon}$ =f($\underline{\delta\sigma},\underline{\sigma}$), where f is a function which depends on the sample story. Moreover, owing to the existence of the quasi-static regime which states that the response $\underline{\delta\varepsilon}$ of the material is independent of the speed of loading (if this one is slow enough), it can be shown that f shall be a homogeneous function of degree 1 in $\underline{\delta\sigma}$ (i.e. f($\lambda \underline{\delta\sigma},\underline{\sigma}$)=$\lambda$f($\underline{\delta\sigma}, \underline{\sigma}$)) [2] ; this means that the response to an increment of stress $\underline{\delta\sigma}$ in a given direction shall vary linearly with its module $||\delta\sigma||$. However, this incremental law can not be strictly linear as a function of $\underline{\delta\sigma}$ in the whole domain of possible increments $\underline{\delta\sigma}$, because such an hypothesis would generate perfect reversibility and would be in contradiction with the well known fact that the evolution of a granular material is not reversible.

For instance, labelling $\underline{\delta\varepsilon}$ the response to the increment of stress $\underline{\delta\sigma}$, and $\underline{\delta\varepsilon'}$ the deformation corresponding to -$\underline{\delta\sigma}$ one shall not have $\underline{\delta\varepsilon'}$ = -$\underline{\delta\varepsilon}$ but shall have $\underline{\delta\varepsilon'} \neq -\underline{\delta\varepsilon}$ [2] . This is why hypo-elastic law is commonly used to model the rheological behaviour of this medium [2,5] .

One of the simplest modelling consists in separating the space E of variation of $\delta\underline{\sigma}$ into few separate subspaces $E_k$ (with $E=\cup E_k$ ) where the rheological law $\delta\underline{\varepsilon}=f_k(\delta\underline{\sigma},\underline{\sigma})$ is strictly linear within each domain in the limit of a $||\delta\underline{\sigma}||$ infinitely small. As it is currently observed experimentally that the response to a strain increment is continuous; we will assume such a property here too, which imposes some relation between the $f_k$ : the responses $f_k(\delta\underline{\sigma},\underline{\sigma})$ and $f_{k+1}(\delta\underline{\sigma},\underline{\sigma})$ at the frontier between the two zones $E_k$ and $E_{k+1}$ shall be equal, i.e. $f_k(\delta\underline{\sigma},\underline{\sigma}) = f_{k+1}(\delta\underline{\sigma},\underline{\sigma})$. It is worth noting that this continuity is achieved spontaneously for two opposite directions $\delta\underline{\sigma}$ and $-\delta\underline{\sigma}$, since the crossing occurs at $\delta\underline{\sigma}=0$ and due to the fact that f is homogeneous of degree 1 in $\delta\underline{\sigma}$..

• *Hypoelasticity applied to plasticity theory:* It has been demonstrated [6, 7] that this incremental approach is able to describe systems obeying perfect plasticity theory and/or elasto-plastic one with one or few different plastic mechanisms. In the case of an elasto-plastic system with a single plastic mechanism, the direction of the plastic deformation of the sample is controlled by the direction of the normal to the load surface, and the amplitude of the plastic deformation is controlled by the hardening law so that it depends linearly on $||\delta\underline{\sigma}||$; the total (elastic+plastic) deformation is then the sum of an elastic response and of a plastic yielding in a precise direction, both being proportional to $||\delta\underline{\sigma}||$ . Since projection operators act as linearly independent mechanism and because they can be added linearly, it turns out that a sample obeying to an elasto-plasticity theory with multiple-mechanism law, all being activated, shall obey the incremental modelling with a linear response by zones in the limit of a large-but-finite number of independent plastic mechanisms.

## 2. Undrained ($v=c^{ste}$) compression test: basis

In the rest of the paper, a granular material will be assumed to obey such an incremental description with a set of linear functions $f_k$ , each one being defined for a zone; furthermore, the number of zones will be assumed small enough, so that triaxial tests performed at $\sigma_2=\sigma_3=c^{ste}$ and at constant mean stress p (i.e. $3p= \sigma_{1o}+ \sigma_{2o}+ \sigma_{3o}=c^{ste}$) pertain to the same linear domain.

We will consider the evolution of a granular sample submitted to an axial compression test by increasing the vertical stress $\sigma_v$ from $\sigma_o$ for which the volume is kept constant (i.e. v= constant). It is called an undrained test in soil mechanics. The initial state of the sample will be assumed isotropic and the initial stress too (i.e. $\sigma_{1o} = \sigma_{2o} = \sigma_{3o} = \sigma_o$). At last, we will consider that principal-stress and principal-strain directions remain parallel to one another all along the test due to the symmetry of the system.

Also, due to the symmetry of the initial state (isotropic stress, isotropic material), the incremental response of this state shall be characterised by two zones (one for $||\delta\underline{\sigma}||>0$, the other for $||\delta\underline{\sigma}||<0$ ), each one is characterised by two independent parameters , i.e. a pseudo Young modulus $C_o$ and a pseudo Poisson coefficient ν. So, in a given

zone, one can write:

$$\begin{pmatrix} de_1 \\ de_2 \\ de_3 \end{pmatrix} = -C_o \begin{pmatrix} 1 & -n & -n \\ -n & 1 & -n \\ -n & -n & 1 \end{pmatrix} \begin{pmatrix} ds_1 \\ ds_2 \\ ds_3 \end{pmatrix} \qquad (2)$$

As paths which are considered here concern only those ones pertaining to a single domain $E_k$, and are concerned with an increase of $\sigma_1$, we will assume that there is only one set of pseudo Young modulus $1/C_o$ and pseudo Poisson coefficient $\nu$ of interest. This set can be determined from triaxial test curves performed at $\sigma_2=\sigma_3=c^{ste}$, since the slope of the curve $\sigma_1$ vs. $\varepsilon_1$ is just $1/C_o$ and since $\nu$ is related to the dilatancy $K = -\delta\varepsilon_v/\delta\varepsilon_1 = -(\partial\varepsilon_v/\partial\varepsilon_1)_{\sigma_2=\sigma_3=cste}$ for a test performed at $\sigma_2=\sigma_3=c^{ste}$ by:

$$K = 2\nu - 1 \qquad (3)$$

When combined with Rowe's relation (Eq.1), Eq. (3) leads to the evolution of the pseudo Poisson coefficient when the sample remains isotropic:

$$2\nu = (\sigma_1/\sigma_3)/(1+M) \qquad (4)$$

## 3. Prediction of an undrained compression if the sample remains isotropic

Applying the hypothesis of undrained compression ($\delta v = \delta\varepsilon_1 + \delta\varepsilon_2 + \delta\varepsilon_3 = 0$) in Eq. (2) imposes :

$$\delta v = 3\delta p (1-2\nu) = 0 \qquad (5)$$

with $\delta p = (\delta\sigma_1 + \delta\sigma_2 + \delta\sigma_3)/3$. This imposes then that either $\delta p = 0$ or $\nu = ½$. If one defines $q = \sigma_1 - \sigma_3$ as it is done in usual, the second condition ($\nu = ½$) is only possible when $q = M'p = M\sigma_3$, with $M' = 3M/(3+M)$ due to Eq. (4). Generalising the notion introduced by Luong [8], we will call this set of states the characteristic states ; they are defined by $q = M'p = M\sigma_3$ so that this set is a plane surface in the (q,p,v) phase space of soil mechanics. So starting from an isotropic stress ($\sigma_1 = \sigma_2 = \sigma_3$), the only solution is that p remains constant till the sample reaches the q=Mp line and the trajectory starts as a vertical segment in the (q,p) plane till it reaches the q=M'p line. At this stage what does it occur?

*The characteristic states (i.e. $q = M\sigma_3 = M'p$ line) are the location of a transcritical bifurcation*
When this q=M'p line is reached, the sample can evolve within two independent ways: either it can pursue its evolution keeping p constant, or it can follow the q=M'p line; the real trajectory the sample shall follow is this one which dissipates the less energy δW. This holds true as far as the experimenter does not bias the experiment by controlling boundary conditions, and as far as the system evolves according to a principle of minimum energy dissipation. It seems that this last principle still remain to be demonstrated in the present case because the two paths are quite distinct so that one cannot applied a variational study. Nevertheless, we will assume that this principle is satisfied.

In this case, the work can be written as $\delta W = q\delta\varepsilon_1 + p\delta\varepsilon_v$, $= C_o\{\delta\sigma_1^2 + 2(1-\nu)\delta\sigma_2^2 - 4\nu\delta\sigma_1\delta\sigma_2\}$ and can be evaluated using Eq. (2) together with the following conditions:

$\delta\sigma_1 = -2\delta\sigma_2 = -2\delta\sigma_3$ , $\delta v = 0$      for the $c^{ste}$-pressure path    (6a)
$\delta q = M'\delta p$ , $\delta\sigma_1 = (1+M)\delta\sigma_2 = (1+M)\delta\sigma_3$     for the $\nu = \frac{1}{2}$ path.    (6b)

After calculation, it is found that the second path dissipates less energy than the first one since their difference $\delta W_{p=cste} - \delta W_{characteristic\ state}$ is positive; for instance, when M is 2 about and $\nu \approx \frac{1}{2}$, one gets:

$$\delta W_{p=cste} - \delta W_{characteristic\ state} \approx (5C/18)(1+13\nu)\delta\sigma_1^2 \tag{7}$$

So, it is positive . Furthermore, from Eq. (6) one gets that the two possible solution for $\delta\sigma_2$ when working at $\delta\sigma_1$ imposed is of the same order of magnitude but of opposite sign ($\delta\sigma_2 = -\frac{1}{2}\delta\sigma_1$ or $\delta\sigma_2 = \delta\sigma_1/(1+M)$).

So the trajectory changes of direction when the sample reaches the q=M'p line; it means that the trajectory in the (q,p) plane turns suddenly on the right on the q=M'p line when reaching this line and after having a vertical line (p=$c^{ste}$) . This is shown on Fig. 1. The trajectory stops at the critical point after a while.

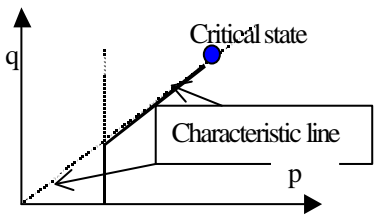 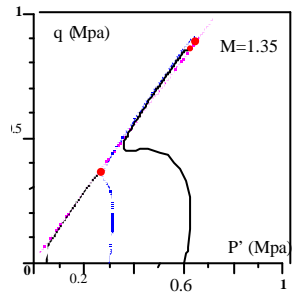

**Figure 1:** predicted trajectory of an undrained triaxial compression on an isotropic sample.

**Figure 2:** Typical experimental results on Hostun sand (after Flavigny) for 3 different pressures and densities.

We report on Fig. (2) typical experimental data obtained on sand. Indeed the prediction (Fig. 1) compares well with experimental data (Fig. 2). The denser the pile and the lower the working pressure, the better the prediction. This is due to the fact that the denser the material, the harder it is so that the smaller the axial and radial deformations, and the longer the contact distribution of the grains remains isotropic; hence the longer the response remains isotropic.

*Few remarks*

• *Transcritical bifurcation* : When reaching the characteristic line the system undergoes a transcritical bifurcation, as it is called in the bifurcation theory and/or in the theory of dynamical systems [10].

• *Initial anisotropy* : on Fig. 2 one can observe that the right-most curve does not start perfectly vertically but that it is slightly inclined; we interpret this feature as the result of a slight initial anisotropy which has been generated during the sample preparation. (see next section for a discussion on anisotropy).

• *Characteristic states* : this notion has been introduced by Luong [8] to characterise a peculiar point of the sample evolution during an axisymmetric compression at $\sigma_2=\sigma_3=c^{ste}$ ; this state is characterised by the absence of contractancy and of dilatancy and by the same stress ratio as the critical state one $q=M\sigma_3$ [11-13]. This notion of characteristic state has been applied successfully to interpret the contractant-dilatant behaviour of soils under cyclic conditions [8-14] and to liquefaction [8-14]. However, to the best of my knowledge, it has never been used associated with an incremental formulation, for which it takes its whole meaning.  For instance, introducing the characteristic state as the state characterised by a pseudo Poisson coefficient equal to ½ and telling that this state is also characterised by a stress field obeying to $q=M\sigma_3$ leads at once to understand why this state does not change of volume during cyclic experiments. Furthermore assuming that the Rowe's relation define the pseudo Poisson coefficient allows to understand at once that a state characterised by a stress field $q<M\sigma_3$ is always contractant and a state for which $q>M\sigma_3$ is always dilatant under cyclic conditions [15].

This is why we consider that the notion of characteristic state is quite well funded and has to be specified in order to correspond to our own definition developed above. We will see in the next section that **this new definition of the characteristic state can be generalised and holds true even when the system does not remain isotropic**.

• *the critical state is the end of the trajectory* : As mentioned already, our modelling predicts that the trajectory shall stop evolving when reaching the critical state, because it is the boundary of the characteristic-state domain.

• *development of anisotropy* : one can observe on Fig. (2) that the vertical trajectory turns left after a while; this occurs for the loser sample and/or for the higher $\sigma_3$ pressure. This can be taken into account by introducing some anisotropy of the response which develops when increasing the axial load and the deformation as we will show now.

## 4. Undrained compression test on anisotropic material:

In this case, Eq. (2) does not hold any more and one has to replace it by a more complex behaviour defined by the relation:

$$\begin{pmatrix} d\varepsilon_1 \\ d\varepsilon_2 \\ d\varepsilon_3 \end{pmatrix} = -C_o \begin{pmatrix} 1 & -\nu' & -\nu' \\ -\nu & \alpha & -\nu'' \\ -\nu & -\nu'' & \alpha \end{pmatrix} \begin{pmatrix} d\sigma_1 \\ d\sigma_2 \\ d\sigma_3 \end{pmatrix} \quad (8)$$

However, due to energy consideration, Eq. (8) can be simplified since one shall have :

$\nu' = \nu$ \quad (9)

(This is obtained by calculating the work $\delta W = \sigma_1 \delta\varepsilon_1 + \sigma_2 \delta\varepsilon_2 + \sigma_3 \delta\varepsilon_3$ for a deformation which is the combination of two infinitesimal deformations in two different directions but pertaining to the same linear zone; writing that this work shall not depend on the order of the two increment implies Eq. (9)).

So we have to evaluate the three parameters $\alpha$, $\nu$, $\nu''$. This can be done using experimental data of Fig. (2). We start with the final part of the trajectory, i.e. the characteristic state segment.

• *End of the trajectory, the anisotropic characteristic state*: As the end of the trajectories remains on the same line q=M'p, one shall conclude that this line does not depend on the anisotropy and that the M' value corresponds also to the one of the critical-state. This allows to conclude that applying a stress increment ($\delta\sigma_1 \neq 0$, $\delta\sigma_2 = \delta\sigma_3 = 0$) at any stage of the deformation when the system has reached the q=M'p line will lead to no volume change. This implies that the q=M'p line is characterised by:

$\nu = \nu' = \frac{1}{2}$ \quad anisotropic characteristic state \quad (10a)

Taking into account Eq. (10), and investigating the undrained path for which ($\delta\sigma_1 \neq 0$, $\delta\sigma_2 = \delta\sigma_3 \neq 0$) and $\delta v = 0$, leads to the condition [16] :

$\alpha - \nu'' = \frac{1}{2}$ \quad anisotropic characteristic state \quad (10b)

• *Beginning of the trajectory when starting from an isotropic sample*: Here we neglect the small anisotropy observed in Fig. (2) at the origin. We remark also that the development of the anisotropy influences the trajectory in the (q,p) plane as a second order effect in term of stress since it is the increase of deviatoric stress which induces the anisotropic deformation of the sample, which generates in turn the anisotropy of the contact distribution and the anisotropy of the mechanical response. So the sample obeys to Eq. (2) at the origin. This condition combined with the Rowe's equation (Eq. 1)) leads to:

$\alpha = 1$ \quad isotropic original state \quad (11a)

$\nu = \nu' = \nu'' = 1/(2+2M)$ \quad isotropic original state \quad (11b)

We remark by passing that Eq. (11b) leads to an initial pseudo Poisson coefficient of 0.17 about, since typical experimental value of M is 2 about, (which corresponds to a typical friction angle of 30°). This initial pseudo Poisson coefficient is not far from those ones observed experimentally quite often. This strengthens our modelling and its incremental formulation.

So this analysis predicts that the trajectory starts vertically in the (q,p) plane. Then it deviates from this line and reaches the q=M'p line. As the Rowe's relation is valid for a triaxial test at $\sigma_2=\sigma_3=c^{ste}$, this fixes the evolution of the pseudo Poisson coefficient $\nu=\nu'$:

$\quad\nu=\nu'= \sigma_1/\{2\sigma_3(1+M)\}$          anisotropic evolution      (12)

In the same way, writing the undrained condition in terms of $\delta q$ and $\delta p$ imposes:

$\quad \delta v=0=C_o\{(2/3)\delta q(1-\nu-\alpha+\nu'')+\delta p(1+2(\alpha-\nu'')-4\nu\}$      (13)

If $C_o$ is non zero, which holds true always in the case of undrained tests (if one excepts the final critical state) since the system resists always according to the classical rheological behaviour of granular matter. So, Eq. (13) leads to the determination of the parameter $\alpha-\nu''$ from a best fit with experimental data and using Eq. (12):

$\quad (\alpha-\nu'')=\{(4\nu-1)\delta p+2(\nu-1)\delta q/3\}/(2\delta p-2/3\delta q)$      (14)

or

$\quad (\alpha-\nu'')=\{(4\nu-1)+[2(\nu-1)/3](\delta q/\delta p)\}/\{2-(2/3)(\delta q/\delta p)\}$      (14)

As $\delta q/\delta p$ is the slope of the tangent to the experimental curve of Fig. (2), the experimental evolution $(\alpha-\nu'')$ can be found directly indeed from the experimental trajectory since the evolution of $\nu$ is known from Eq. (12) and since $\sigma_1$ and $\sigma_3$ are given by:

$\quad \sigma_1=(2q+p)/3$                                            (15a)
$\quad \sigma_3=p-q/3.$                                             (15b)

## 5. Discussion:
### *work function, plastic behaviour and perfect plasticity*

In a previous paper [17], Stefani and the present author have proposed a simple modelling of the granular-material rheology. It assumes that the work dissipated in the material during a deformation can be written as:

$\quad \delta W = f(q,K,p)\delta\epsilon_1$      (16)

where f is a function which depends explicitly on the 3 parameters q, p and the dilatancy K. As this dissipated energy shall be equal to the work given to the sample from outside, this presentation allowed to predict few characteristics of the rheology of granular materials (these ones are: uniqueness of dilatancy under isotropic stress, existence of a unique well defined stress ratio q/p for the characteristic state (K=0) and for the critical one, coincidence between the maximum of q and the maximum of dilatancy...[17]). Furthermore we argued at that time i) that this presentation might be equivalent to the one proposed by Schofield & Wroth [12] if one assumed the right function f and ii) that it led to the Rowe's relation if we used another function f [18].
At this stage a question arises: is Eq. (16) in agreement or in contradiction with a perfect-plasticity modelling? The answer can be done in few steps:
i) The fact that f depends explicitly on K seems to assume that the way the system deforms is an internal variable which shall be found for a given process; so, this seems to be in complete contradiction with the perfect-plasticity theory which assumes on the contrary that the way the system deforms is fixed by the stress field.
ii) Hence, writing an equation such as Eq. (16) imposes to use an incremental formalism instead of a perfect plasticity modelling.
iii) But in turn, writing that the deformation path is imposed by Eq. (16) has no sense in an incremental modelling, since this deformation depends on the direction of the increment of stress.
iv) Thus we believe that writing an equation equivalent to Eq. (16) is just simply to assume an incremental modelling with a variable pseudo Poisson coefficient. This is just what we have done all along this paper and in ref. [3].

*Validity of the mean field treatment:*
We have explained in the introduction how Rowe derived his relation concerning the dilatancy mechanism. The generalisation he used is a typical mean field reasoning. As it compares well with experimental data, it seems that his argumentation is validate by them. So, at first sight, one is led to conclude that the perfect-plasticity theory is the adequate way of thinking when using the Rowe's law. However, this paper demonstrates just the contrary and it shows the efficiency of the incremental modelling. Thus, what is wrong in Rowe's demonstration?

The answer to this question is: what is wrong is the compatibility of men field approach with the perfect plasticity hypothesis, since both are incompatible to each other, because the former assumes a set of possible deformation while the later a single deformation path.

So, it is worth ending this paper stressing the danger of a mean field treatment: in general it is a powerful technique which leads to quite good approximate results, when well used. However, the assumptions which are made are not so well controlled most often so that it may lead to erroneous thinking which may lead to a large waste of time. We believe that it was happening in the case of granular material mechanics [19]. But we hope that it will disappear soon.

## 6. Conclusion:

In a previous paper, we have demonstrated the efficiency of the incremental modelling associated to the Rowe's law of dilatancy to understand the rheological behaviour of granular material under oedometric test; this has allowed to derive theoretically the $K_o$ value of the earth pressure ratio at rest and to demonstrate that it corresponds approximately to the best fit obtained by Jaky ($K_{jaky}=1-\sin\varphi$). Here we have extended this formulation and applied it to undrained tests. We have found that the trajectory of the stress field start with constant pressure and ends along the line of characteristic states. These features are observed experimentally.

Furthermore, this modelling has allowed to get a deeper understanding of the nature of the characteristic states so that we have extended and specified the definition of these states, especially when anisotropy is developed. We have also shown that the trajectory of the sample under undrained compression undergoes a trans-critical bifurcation when it reaches the characteristic states from an initial isotropic stress.

It is remarkable that the main predicted features are observed experimentally indeed. This seems to strengthen the concepts used and the experimental results on which these concepts were defined; it shows also the coherence of experimental data.

At last, it is worth noting the interest of this formulation since i) it is quite simple, ii) it is in agreement with most of the previous plastic formulation, iii) it is in agreement with previous well accepted rheological features of soils and granular materials and iv) it gets a simple way to explain most of the features observed experimentally. So the investigation of this formulation has to be pursued.

At last, the validity of this approach means likely that the derivation proposed by Rowe of the Rowe's equation may be rather fortuitous since it is based on a single plastic mechanism .

**References**

[1] P.W. Rowe, "The stress dilatancy relation for static equilibrium of an assembly of particles in contact", *Proc. Roy. Soc. Lndn* **A269**, 500-527, (1962)

[2] F. Darve, "L'Ecriture incrémentale des lois rhéologiques et les grandes classes de lois de comportement", *Manuel de Rhéologie des géomatériaux*, ed. F. Darve, presses des Ponts & Chaussées, Paris, 129-152, (1987)

[3] P. Evesque, "On Jaky Constant of Oedometers, Rowe's Relation and Incremental Modelling" , *poudres & grains* **6**, 1-9, (1999); P. Evesque, (1997), "Stress in static sand piles: Role of the deformation in the case of silos and oedometers", *J. de Phys. I France* **7**, 1501-1512, (1997)

[4] E. Frossard, "Une équation d'écoulement simple pour les matériaux granulaires", *Géotechnique* **33**, 21-29, (1983)

[5] J. Tejchman, (1997), "*Modelling of shear localisation and autogeneous dynamic effects in granular bodies*", PhD Thesis, Karlsruhe

[6] B. Loret, (1987), Elastoplasticité à simple potentiel, *Manuel de Rhéologie des Géomatériaux*, Presses de l'ENPC, Paris, pp 157-187

[7] B. Loret, (1987), Application de la théorie des multimécanismes à l'étude du comportement des sols, *Manuel de Rhéologie des Géomatériaux*, Presses de l'ENPC, Paris, pp. 189-214

[8] M.P. Luong, *C.R. Acad. Sc. Paris*, 287, Série II, pp. 305-7 , (1978)

[9] This assumption is not obvious since the system has to choose between two distinct definite and completely different ways. So it is not as if all possibilities between these two paths were possible, for which one could have applied some variational hypothesis to solve the dynamics. So, this assumption remains to be demonstrated in the present case.

[10] P. Manneville, P. Manneville, *Structures dissipatives, chaos et turbulence*, édition Aléa, Saclay, Commissariat à l'Energie Atomique, (1991) ; *Dissipative structures and weak Turbulence*, Accademic press, (1990)

[11] Roscoe K.H., Schofield A.N. & Wroth C.P., 1968, On the yielding of soil, *Geotechnique* **VIII**, 22-53

[12] Schofield A.N. & Wroth C.P., 1968, Critical State of Soil mechanics, Pergammon press

[13] J.H. Atkinson & P.L. Bransby, *The Mechanics of soils*, Mac Graw Hill, Cambridge Un. Press (1977)

[14] A. Pecker, *Dynamique des sols*, Preses des Ponts et Chaussées, Paris (1984)

[15] We assume implicitly here that the system is governed by two incremental zones, one corresponds to loading and to irreversible effects, the other one to unloading and to elastic response; the loading zone is the zone which obeys the Rowe's equation.

[16] We remark by passing that the anisotropic characteristic state is characterised by the relation $1-v=\alpha-v''$ which was needed to be satisfied in order to get a value of the oedometer stress ratio independent of the anisotropy. So, this result seems to strengthen the results of papers [3]. This is only at first sight however, since applying the same relation $1-v=\alpha-v''$ everywhere leads to predict that the trajectory of an undrained test remains vertical till it reaches the $q=M'p$ line; hence this disagrees with what is observed in Fig. (2) when anisotropic response develops itself.

[17] Evesque P. & C. Stéfani, *J. de Physique II France* **1**, 1337-47 (1991); *C.R. Acad. Sci. Paris* **312**, série II, 581-84 (1991)

[18] An approach similar to the one of [17] was proposed in [13] but it did not lead to all results predicted in [17].

[19] For instance, it is worth noting that papers of ref. [3] which establish the efficiency of the incremental modelling combined with the Rowe's law was rejected by mechanics specialists of the *Compte rendus de l'Accadémie des Sciences* in 1997; at that time, the main reason given by the two referees was the incompatibility between with the proposed incremental modelling and the Rowe's approach; in the same way, the version published in *Poudres & Grains* was rejected in 1999 by *Geotechnique*. On the other hand, the tenants of the incremental modelling did not consider the Rowe's law in well consideration (Gudehus private letter).

The electronic arXiv.org version of this paper has been settled during a stay at the Kavli Institute of Theoretical Physics of the University of California at Santa Barbara (KITP-UCSB), in june 2005, supported in part by the National Science Fundation under Grant n° PHY99-07949.

*Poudres & Grains* can be found at :
http://www.mssmat.ecp.fr/rubrique.php3?id_rubrique=402